# Multi-proton bunch driven hollow plasma wakefield acceleration in the nonlinear regime

**Authors:**
Yangmei Li[1,2,a)], Guoxing Xia[1,2], Konstantin V. Lotov[3,4], Alexander P. Sosedkin[3,4], Kieran Hanahoe[1,2], Oznur Mete-Apsimon[1,2]

**Affiliations:**
[1]School of Physics and Astronomy, University of Manchester, Manchester M13 9PL, UK.
[2]Cockcroft Institute, Warrington WA4 4AD, UK.
[3]Budker Institute of Nuclear Physics, Novosibirsk 630090, Russia.
[4]Novosibirsk State University, Novosibirsk 630090, Russia.

**ABSTRACT**

Proton-driven plasma wakefield acceleration has been demonstrated in simulations to be capable of accelerating particles to the energy frontier in a single stage, but its potential is hindered by the fact that currently available proton bunches are orders of magnitude longer than the plasma wavelength. Fortunately, proton micro-bunching allows driving plasma waves resonantly. In this paper, we propose using a hollow plasma channel for multiple proton bunch driven plasma wakefield acceleration and demonstrate that it enables the operation in the nonlinear regime and resonant excitation of strong plasma waves. This new regime also involves beneficial features of hollow channels for the accelerated beam (such as emittance preservation and uniform accelerating field) and long buckets of stable deceleration for the drive beam. The regime is attained at a proper ratio among plasma skin depth, driver radius, hollow channel radius, and micro-bunch period.

# I. INTRODUCTION

Plasma wakefield acceleration (PWFA)[1] has attracted much attention due to ultra-high accelerating gradients sustained by plasmas in comparison with those in metal cavities of radio-frequency accelerators. Nevertheless, the obtainable energy gain of the witness bunch in one acceleration stage is subject to the transformer ratio limit[2] and thereby cannot significantly exceed the energy of driver particles. Under this circumstance, proton driven PWFA[3-8] (PD-PWFA) looks particularly competitive as existing proton bunches have huge energy and large populations. For instance, the nominal bunches in the Large Hadron Collider (LHC) are of 6.5 TeV in energy and of $1.15\times10^{11}$ protons in bunch population, which corresponds to 125 kJ/bunch. It significantly exceeds the electron or positron bunch energy created at the Stanford Linear Accelerator Center[9] (0.12 kJ/bunch) or the proposed International Linear Collider (0.8 kJ/bunch). The energy of laser pulses is even less. For instance, the highest energy of a PW laser proposed to be used for wakefield acceleration is about 40 J[10]. Owing to available high energy contents, proton bunches become very promising candidates in the near future, which are capable of bringing witness particles to the energy frontier in a single plasma stage.

Currently available high-energy proton bunches are tens of cm long, which is much longer than the usual submillimeter plasma wavelength. Such long proton bunches therefore hardly excite strong plasma wakefields directly. Longitudinal compression of the bunch length to the plasma wavelength by traditional methods is conceivable[11-14], but it will be too costly and technically challenging to implement given the required compression factor of up to several orders of magnitude. Fortunately, the self-modulation instability (SMI)[5,14-16] offers a new way to cope with this issue. The SMI is the seeded axisymmetric mode of the transverse two-stream instability[17,18] that develops in the beam-plasma system. The transverse wakefield of the beam leads to rippling of the beam itself, which further amplifies the plasma waves. Because of this positive feedback, the long proton bunch transforms into a train of micro-bunches that follow equidistantly at the plasma wavelength[19-21]. With proper seeding, the axisymmetric mode develops quickly and suppresses destructible non-axisymmetric modes like the hosing instability[22-24]. As a result, the beam resonantly excites plasma waves with large amplitudes. The AWAKE experiment at CERN[25-28] follows this concept and is the world's first proof-of-principle experiment of long proton bunch driven PWFA. Several related experiments with long or pre-bunched electron beams[29-35] are proposed or conducted to enhance the understandings of the SMI and resonantly driven waves.

While the SMI forces the micro-bunching of the long proton bunch and allows harnessing high energy protons in the PWFA, it also limits the beam-plasma interaction to the linear or weakly nonlinear regimes[5,36,37]. This is because the wave period elongates as nonlinear effects comes into play, and the wave eventually falls out of resonance with the bunch train[36]. The plasma focusing in the linear regime depends on the witness density and, for a high witness charge, nonlinearly varies both across and along the witness bunch[38]. The accelerating field also inevitably varies radially instead of being constant, which is detrimental to the witness quality[39].

Hollow plasma channels[40] are one of two solutions to the deterioration issues of witness quality currently discussed in the context. Another solution is the strongly nonlinear (blowout) regime in which the witness bunch propagates in a "bubble" void of plasma electrons, but containing plasma ions[41,42]. The blowout regime is easier to implement experimentally, thus it has been extensively studied during the last two decades[43,44]. However, it has several potential drawbacks. Very dense particle beams required for collider applications can produce large perturbations in the ion density, giving rise to nonlinear transverse fields inside the bubble and degradation of the witness quality[45]. Massive ions with heavier elements are less prone to density perturbations, but incur a prohibitively large growth of witness emittance from multiple Coulomb scattering[46,47]. Furthermore, the bubble naturally appears for laser or electron drivers, but not for protons or positrons. There is a blowout-like regime for positively charged drivers[3,4,48], but the witness bunch quality degrades more due to the presence of plasma electrons in the bubble. Last but not least, the blowout regime is asymmetric with respect to the witness charge, and the accelerating bucket for positrons is extremely narrow[49]. Hollow channels are free from these drawbacks, and this stimulates theoretical studies even though experimental implementation of such channels is still in its infancy[50].

Up to now, all theoretical studies of hollow channels were dedicated to short drivers (both for particle beams[6-8,51-54] and for lasers[40,55-62]). In particular, an advantageous nonlinear regime was discovered for short proton bunches of a high charge[6-8]. However, these bunches are difficult to produce. In this paper, we demonstrate that equally good acceleration conditions are possible with trains of short, lower charge proton bunches. No strong dissipation of the accelerating mode is observed, which enables high efficiency of multi-bunch wave excitation. We simulate the beam dynamics over a long propagation distance and observe the preservation of the witness beam quality. We also discover that, because of no defocusing from background ions in the channel, the regions of stable deceleration for protons (deceleration buckets) are longer than in the uniform plasma and reach almost half of the wave period.

Here is the arrangement of this paper. In Sec. II, we introduce the simulated case that we use for illustrating the multi-bunch driven acceleration regime. Then we elucidate the characteristics of nonlinear wakefields driven in hollow plasma and the behavior of multiple proton bunches in the wakefield potential wells in Sec. III. After that in Sec. IV we demonstrate the acceleration characteristics and conservation of beam quality of the witness bunch, followed by the discussion in Sec. V regarding result dependence on four important quantities: plasma skin depth, driver radius, hollow channel radius and micro-bunch period. We draw conclusions in Sec. VI.

## II. BEAM AND PLASMA PARAMETERS FOR SIMULATIONS

In our proposed scheme, ten identical and equidistant proton bunches enter an axisymmetric hollow channel of radius $r_c$. The surrounding plasma outside the channel is of uniform density $n_p$. The bunches initially have the density distribution

$$n_b = \frac{N_b}{2\sigma_r^2 \sigma_z (2\pi)^{3/2}} e^{\frac{-r^2}{2\sigma_r^2}} [1 + cos\left(\sqrt{\frac{\pi}{2}} \frac{(z-z_i)}{\sigma_z}\right)], |z - z_i| < \sqrt{2\pi}\sigma_z$$

where $\sigma_r$ and $\sigma_z$ are the bunch radius and length, $N_b$ is the bunch population, and $z_i$ is the centroid of the $i$-th bunch. The shifted cosine shape is a convenient approximation to the Gaussian shape with the root-mean-square length of $\sigma_z$, but without infinitely tails. The bunch train period is $\approx 10\sigma_z$. The electron witness bunch has a similar shifted cosine shape and is initially positioned beyond the maximum accelerating field to extend the acceleration distance[8]. The witness charge and shape are not optimized for the minimum energy spread, as we do not intend to demonstrate the capabilities of the optimal beam loading. The initial witness radius is much smaller than the channel radius, which is necessary for the sake of emittance conservation. Still, it is not as small as required for collider applications, since this would require reducing the simulation grid size and increasing the simulation time accordingly.

In order to resonantly drive strong plasma waves, the bunches are supposed to be in tune with the wakefields and reside in the decelerating phases. Presence of the hollow channel results in longer wakefield wavelength than in uniform plasma. Given the bunch period, the wider the channel, the larger the required plasma density. The channel must be wide enough to accommodate most of the driver, so we choose $r_c = 3\sigma_r$ and adjust the plasma density to fit the resonance between the bunches and wakefields. Similarly to Refs. 3, 6 and 8, we surround the plasma by focusing quadrupole magnets, which keep the driver head from emittance-driven erosion and control the witness trajectory.

The beam and plasma parameters are given in detail in Table 1. A typical energy of 1 TeV is chosen as the driver energy. Such high energy also allows us to verify the long term (150 m in our simulated case) bunch propagation and excitation of nonlinear wakes. The simulations are performed with a 2D quasi-static particle-in-cell code LCODE[63,64] using the cylindrical geometry $(r, \theta, \xi)$, where $\xi = z-ct$.

**Table 1: Parameters for simulation**

| Parameters | Values | Units |
|---|---|---|
| **Initial proton beam:** | | |
| Population of a single bunch, $N_b$ | $1.15\times10^{10}$ | |
| Energy, $W_{d0}$ | 1 | TeV |
| Energy spread | 10% | |
| Single bunch length, $\sigma_z$ | 63 | μm |
| Beam radius, $\sigma_r$ | 71 | μm |
| Angular spread | $5\times10^{-5}$ | |
| Bunch train period | 631 | μm |
| **Initial witness electron beam:** | | |
| Population | $2\times10^9$ | |
| Energy, $W$ | 10 | GeV |
| Energy spread, $\delta W/W$ | 1% | |
| Bunch length, $\sigma_{zw}$ | 15 | μm |
| Beam radius, $\sigma_{rw}$ | 10 | μm |
| Normalized emittance, $\varepsilon_n$ | 2 | mm mrad |
| **Unperturbed hollow plasma:** | | |
| Plasma density, $n_p$ | $6\times10^{15}$ | cm$^{-3}$ |
| Hollow channel radius, $r_c$ | 200 | μm |
| **External quadrupole magnets:** | | |
| Magnetic field gradient, $S$ | 0.5 | T/mm |
| Quadrupole period, $L_q$ | 0.9 | m |

## III. NONLINEAR WAKEFIELD EXCITATION IN HOLLOW PLASMA

In the described regime, most driver protons are initially located within the hollow channel radius (Fig. 1a). Protons that start inside the surrounding plasma experience a positive radial force (Fig. 1c) and are quickly defocused (Fig. 1e). This results in some change of the wave drive efficiency in the first few meters of propagation (Fig. 1a). The amplitude of the plasma wave increases with the distance from the driver head. This leads to nonlinear elongation of the wakefield period[36]. Hence the wake is not strictly periodic. Nevertheless, since the elongation is insignificant, all bunches still reside in the decelerating phases and see rising decelerating gradients on average.

In view of the perturbation of plasma electron density (Fig. 1b) and the bunch portraits (Fig. 1e), we see that the first driving bunch stays in a weakly nonlinear wake and the subsequent bunches stay in almost spherical plasma bubbles and experience nonlinear waves of increasing amplitude. The maximum radius of each bubble is larger than the hollow channel radius. It suggests the existence of areas out of the hollow channel but within the plasma electron sheath, where the uniformly distributed plasma ions are more prevalent than the plasma electrons being pulled there. These areas correspond to the yellow and red regions in Fig. 1c denoting positive transverse wakefields. These positive fields will defocus the protons if they get there. The blue area within the channel in Fig. 1c imposes focusing on the protons. It is created by plasma electrons attracted into the channel, whose trajectories are seen in Fig. 1b.

The proton bunches reside in rear halves of the bubbles and are almost half a period long. The bunches completely fill decelerating phases of the wave (Fig. 1a) and nevertheless stably propagate along the channel over a long distance (Fig. 1e), which indicates that the bunches are strongly focused. This contrasts to wave characteristics in uniform plasmas, where regions of simultaneous deceleration and focusing are as short as a quarter of the wave period for low-amplitude waves[2] and even shorter for nonlinear waves and positively charged beams[44,65]. Unique focusing ability of the nonlinear proton-driven wave is attributed to the basin-like radial profile of the wakefield potential in the channel (Fig. 1d, Fig. 2), i.e., the potential drops with decreasing radius within the channel (Fig. 2d). In turn, this particular potential structure appears as protons experience a negative (focusing) or zero radial force all through the channel (Fig. 1c). Regions of strong defocusing force, which are necessary for potential oscillations and longitudinal fields, are thus located outside the channel and have no effect on the beam.

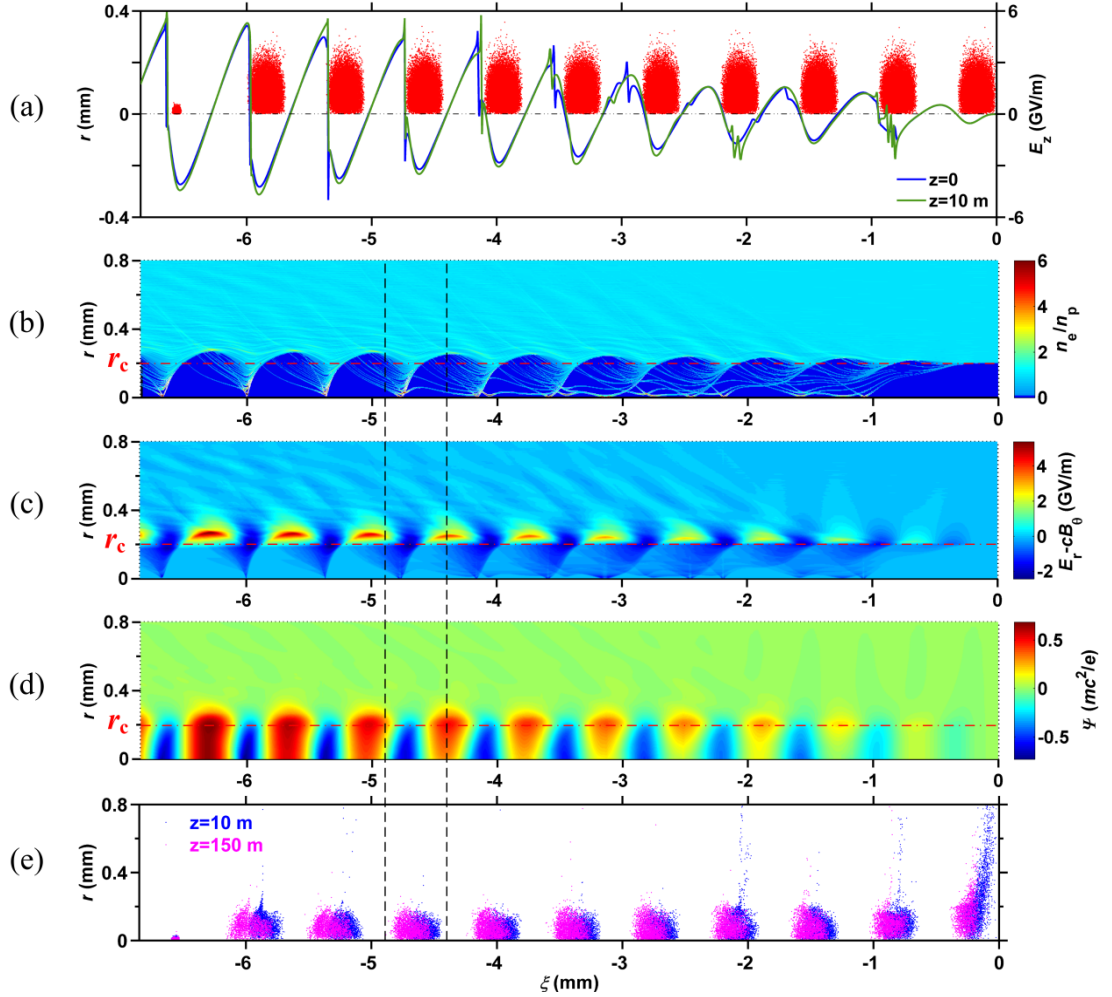

FIG. 1. (a) Initial distribution of 10 proton driving bunches and the witness electron bunch in the real space and the on-axis longitudinal electric fields at $z=0$ and $z=10$ m, respectively. The plasma electron density (b), transverse plasma wakefields (c) and wakefield potential (d) at $z=10$ m. (e) Snapshots of the driver and the witness bunches at $z=10$ m and $z=150$ m, respectively. The red dash-dotted lines in (b), (c) and (d) illustrate the hollow channel boundary. Two vertical black lines are sketched to facilitate the discussion in the following. The observation window travels to the right at the speed of light.

Let
$$W_t = p_r^2/(2\gamma_b) + \Delta\psi(r,\xi)$$
where $p_r^2/(2\gamma_b)$ denotes the transverse kinetic energy of a proton ($p_r$ is its radial momentum and $\gamma_b$ is its relativistic factor), and $\Delta\psi(r,\xi)$ represents the potential energy difference between the value at the proton position and the maximum value along the radius (at the basin edge). If $W_t > 0$, the proton has enough kinetic energy to escape from the potential well. Otherwise it stays in the well and inside the channel. Most of driver protons have $W_t < 0$ all the interaction time (red dots in Fig. 2). More than 90% of protons in total survive over 150 m (Fig. 3). Small proton loss occurs at the boundary areas between positive and negative potentials, where the magnitude of $\Delta\psi(r,\xi)$ is very small due to the lack of plasma electrons inside the channel (like the green line in Fig. 2d). This loss can also be observed in Fig. 1e where a narrow slice of protons escapes out of the hollow channel. Another reason for the observed proton loss is unphysical and related to the final width of the simulation window. There is no plasma focusing at the head of the first bunch, so the protons perform large radial oscillations under the weak focusing of external quadrupoles (Fig. 4a). Some of the protons exit the simulation window radially and are counted as lost in Fig. 3. In our case, protons in the first bunch being excluded this way account for 2.8% of the total driver charge.

Protons at the back of the first bunch experience stronger plasma focusing, thus oscillating with higher betatron frequency and within the hollow channel (Fig. 4b). Fig. 4c gives the trajectory of a proton in the seventh bunch initially located in a deep potential well. The proton oscillates well within 200 μm with much higher betatron frequency.

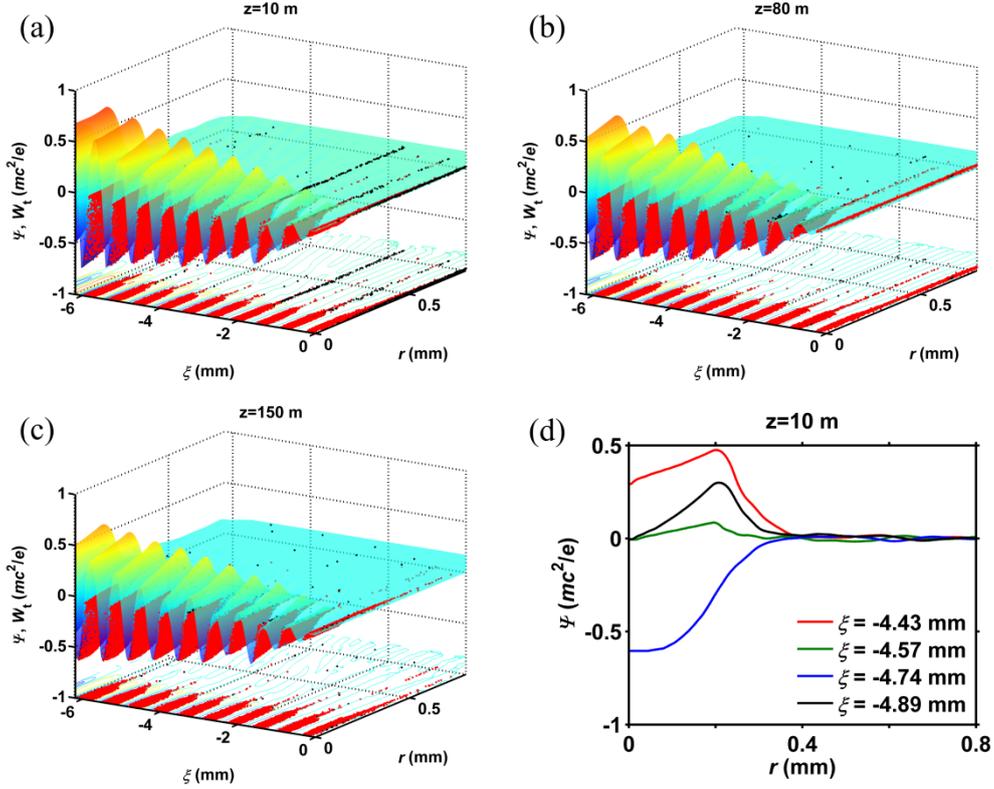

FIG. 2. Spatial distributions (a, b, c) of wakefield potential (surface) and driver protons (red dots for trapped and black for untrapped ones) at different propagation distances. (d) Radial dependencies of the potential at different $\xi$-positions (z=10 m) where the on-axis potential has local extrema or zero, i.e., $\xi$=-4.43 mm (maximum), $\xi$=-4.57 mm (zero), $\xi$=-4.74 mm (minimum), and $\xi$=-4.89 mm (zero). The two vertical black lines in Fig. 1 mark the range $\xi \in$[-4.89 -4.43] discussed here.

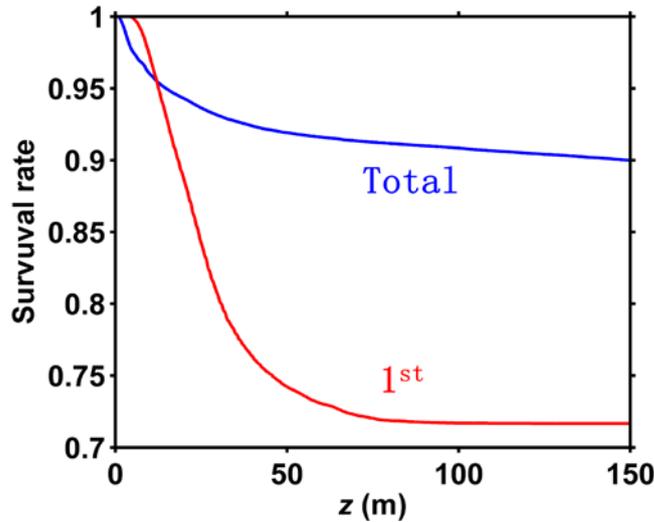

FIG. 3. Survival rates for the whole proton driver and the first driving bunch, respectively.

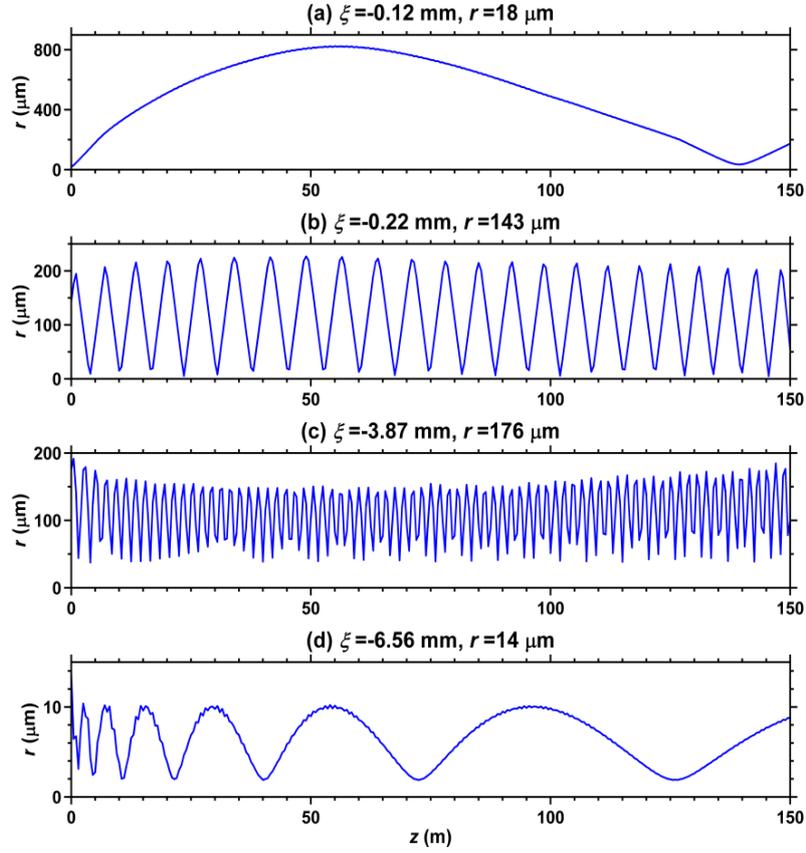

FIG. 4. Trajectories of protons initially located at the head (a) and back (b) of the first driving bunch and at the front (c) of the seventh driving bunch and the trajectory of the witness electron (d).

## IV. ACCELERATION CHARACTERISTICS OF THE WITNESS BUNCH

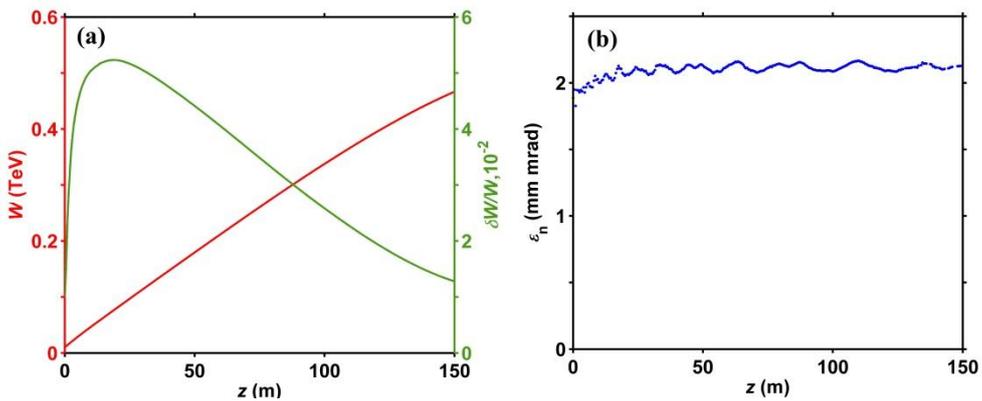

FIG. 5. (a) Energy (red line) and correlated energy spread (green line); (b) normalized emittance (blue points) of the witness bunch with respect to the acceleration distance.

The witness electron bunch is placed in the bubble behind the last proton bunch (Fig. 1). The maximum accelerating gradient in the witness region is 4.4 GV/m, which constitutes 60% of the wave-breaking field for the surrounding plasma. The witness is only focused by the quadrupoles which keep its radius small and prevent it from being deflected by the plasma electrons staying in larger radii. Therefore, in our studied scheme, long intervals of proper focusing for the positively charged driver and for the electron witness coexist. After being accelerated over 150 m, the witness bunch reaches energy of 0.47 TeV with an energy spread as low as 1.3% (Fig. 5a). The normalized emittance of the witness bunch stays at the initial level of 2 mm mrad over this distance (Fig. 5b). After that the witness quality starts to degrade because the depleted driver bunches change their shape, which causes mismatch of the wave and the bunch train, penetration of plasma electrons into the witness region, and defocusing of witness electrons.

The trajectory of a witness electron is shown in Fig. 4d. The electron oscillates within 10 μm, and the oscillation period increases as the energy grows. Since witness electrons only experience weak quadrupole focusing and the amplitude and frequency of betatron oscillations are small, the betatron radiation can be foreseen to be modest.

## V. DISCUSSION

To attain high energy gain and high beam quality of the witness bunch in the proposed scheme, four parameters, i.e., plasma skin depth, driver radius, hollow channel radius and bunch period, need to have proper relative scale factors. Numerical optimization suggests that these parameters should form a ratio of $1:1:3:3\pi$. This result was obtained with simulations and currently lacks a complete theoretical justification as no purely analytical model can predict the outcome as of today. Nevertheless, we present several considerations to back up this result.

First, the driver period must match the wake period to resonantly excite strong wakefields. Given a plasma density, the wake period depends on the channel radius. In our case it is 1.5 times longer than the wavelength in the uniform surrounding plasma. In turn, for a given period of the bunches, more (less) dense plasma should accompany the choice of a larger (smaller) channel radius. For a train of N bunches, the period mismatch should be below $1/N$, which is 10% in our case. Simulations confirm this estimate: channels with radii ranging from 160 μm to 220 μm and surrounding plasma densities between $5\times10^{15}$ cm$^{-3}$ and $7\times10^{15}$ cm$^{-3}$ work almost equally well.

The optimal ratio between the driver radius and the hollow channel radius comes from a compromise between the survival rate of the protons and the accelerating field amplitude. Reducing the channel radius to 160 μm (by 20%) while decreasing the plasma density to keep the resonance leads to the loss of 15% more protons (Fig. 6a). Although the accelerating gradient is initially enlarged by 42% (Fig. 6b), it only lasts for a very short distance and decreases quickly with proton loss. This is because protons in large radii out of the hollow boundary easily escape due to positive radial wakefields. When increasing the channel radius to 240 μm, we obtain a survival rate as high as 95% (Fig. 6a), but the gradient decreases considerably (Fig. 6b).

The optimal ratio between the plasma skin depth and hollow channel radius aims at providing a strong wakefield on the axis, presence of plasma electrons in most cross-sections of the channel, and availability of plasma electron-free bubbles for hosting the witness bunch. This ratio depends on the charge of driver bunches and needs to be adjusted for each particular driver. If the plasma density is lower than the optimal one and the skin-depth is longer (Fig. 7a-c), then there are fewer plasma electrons in the channel, which is unfavorable for driver focusing and subsequently destroys the accelerating field. If the plasma density is higher (Fig. 7d-f), then the bubble structure disappears, bunch contributions stop adding up, and the accelerating field lowers.

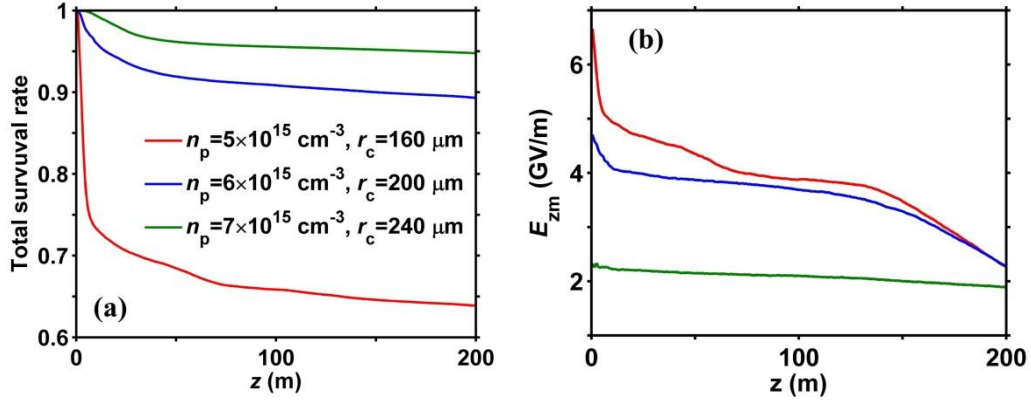

FIG. 6. Driver survival rates (a) and maximum longitudinal wakefield amplitudes (b) for different plasma structures. Each case is denoted by the identically colored lines in (a) and (b). To avoid contributions from field singularities (Fig. 1a) into the wakefield amplitudes, the on-axis electric field $E_z$ is first integrated along $\xi$ and then multiplied by the wake wavenumber and averaged in wide (1 m) intervals in $z$.

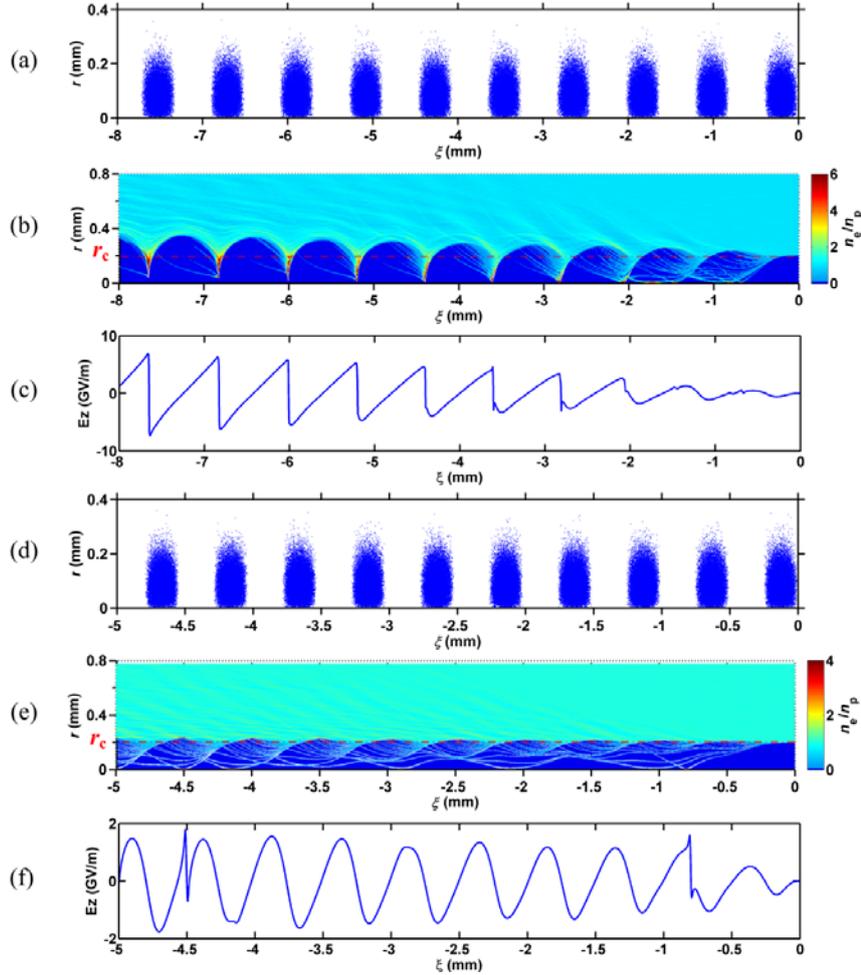

FIG. 7. Initial driver distribution in the real space (a, d), plasma electron density (b, e), and on-axis longitudinal electric field (c, f) for plasma densities of $3\times10^{15}$ cm$^{-3}$ (a, b, c) and $12\times10^{15}$ cm$^{-3}$ (d, e, f). Driver radius, hollow channel radius, and peak driver density are the same as in the baseline variant; the driver period is adjusted to match the wake period. The red dash-dotted lines in (b) and (e) denote the hollow channel boundary.

To get a deeper insight into the effect of channel inhomogeneity, we introduce the perturbation to the density of the surrounding plasma as

$$n = n_0(1 + \delta n \sin(2\pi z/L)),$$

where $n_0 = 6 \times 10^{15}\ cm^{-3}$ is the baseline plasma density, $\delta n$ is the perturbation amplitude and $L$ is the perturbation period. The wakefield characteristics in different cases (L=0.1 m, 1 m and $\delta n$ =0.5%, 2.5%, 5%) are illustrated in Fig. 8. In addition, an irregular density non-uniformity following the form

$$n = n_0(1 + 0.025(1 - cos(\frac{2\pi z}{1\ m})) + 0.015 sin(\frac{2\pi z}{3\ m}) + 0.015 sin(\frac{2\pi z}{21\ m}))$$

is introduced to further validate the proposed scheme. The conclusions drawn from Fig. 8 are fourfold. First of all, the proposed scheme can tolerate a regular density perturbation up to a level of 5% and is slightly more sensitive to the irregular perturbation (Fig. 8b). Second, the oscillations of the maximum longitudinal wakefields follow the perturbation periods (Fig. 8a) and reducing the periods below the proton oscillation period can significantly boost the average wakefield amplitudes (the green line in Fig. 8b). Furthermore, a rise of the plasma density generally damps the wakefields while a drop of the plasma density increases the wakefields (Fig. 8a) similarly to the uniform plasma case[66]. Last but not least, a "clean" accelerating region for the witness beam can still be kept under the plasma density perturbation discussed here and thereby its normalized emittance is well preserved (Fig. 8c). As a consequence, the limitation on tolerable density perturbations comes from decreasing the acceleration rate (Fig. 8d), rather than degrading the witness quality as in the case of uniform plasmas[66].

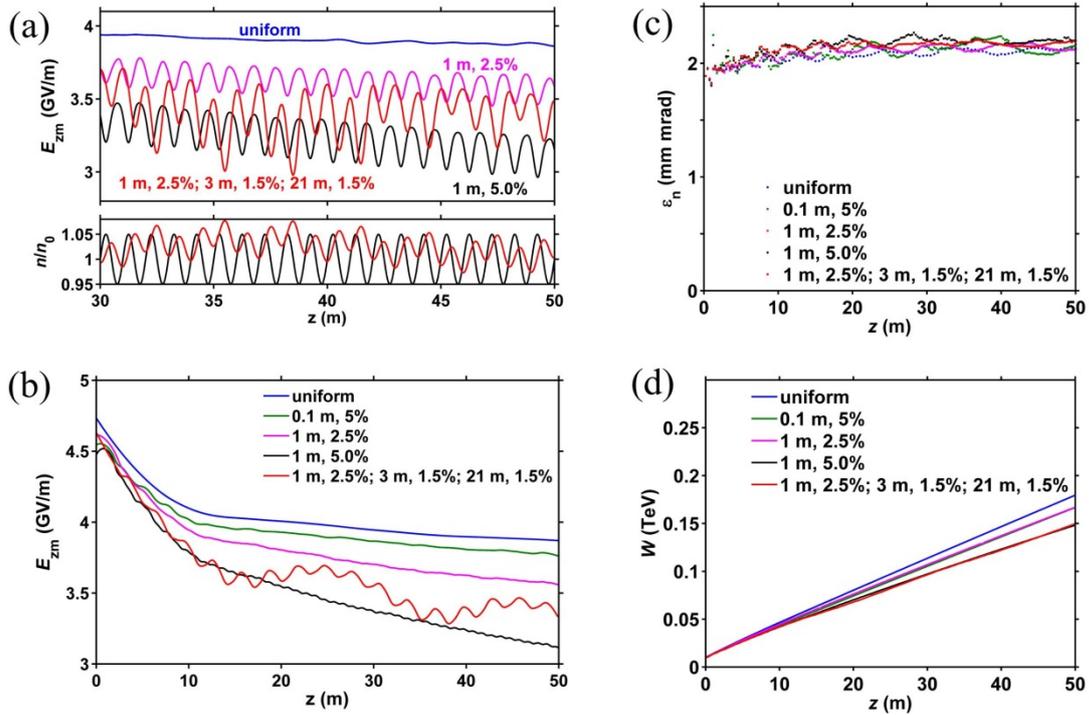

FIG. 8. (a) Maximum longitudinal electric fields excited in axially non-uniform plasmas versus the propagation distance. The field amplitudes (depicted in upper plots) are smoothed in narrow intervals in z so that the oscillations are observable. The lower plots represent the corresponding plasma density profiles. The identically colored lines denote the same cases. (b) Maximum longitudinal electric field amplitudes are averaged in wide (1 m) intervals in z to facilitate the comparison of different cases of plasma inhomogeneity. (c) Normalized emittance of the witness beam in different cases of plasma inhomogeneity. (d) Mean energy of the witness beam versus the acceleration distance. Note that the "uniform" in the legends denotes the hollow channel with axially uniform plasma.

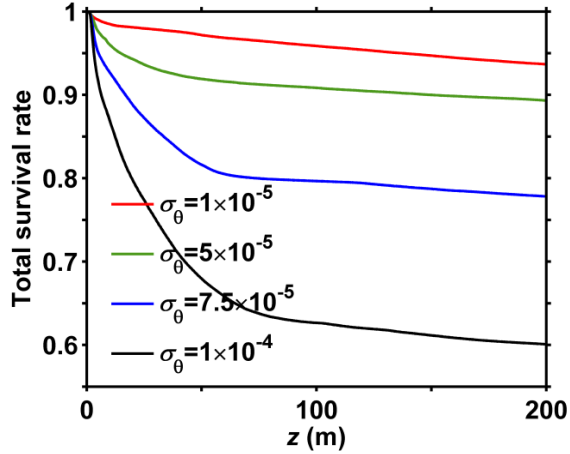

FIG. 9. Total survival rates concerning the whole proton driver with different initial angular spreads.

In Sec. III, we demonstrate that the survival rate of protons is essentially determined by the relation between their radial momenta and the potential well depth. Increasing the angular spread will cause more proton loss, as the protons with larger angular spread have enough transverse kinetic energy to escape from the potential well (Fig. 9). The normalized emittance of the protons in the simulated case (3.5 mm mrad) is typical for state-of-the-art beams[27,67], so there is no safety margin in beam emittance. Therefore, the multiple proton bunches are preferably generated by longitudinal modulation[68] instead of the radial self-modulation[15,16] from a long proton bunch. The reason behind it is that the self-modulation gives rise to large radial momenta of the micro-bunches[21].

## VI. CONCLUSIONS

In summary, by employing hollow plasma we enable the operation of multiple proton bunch driven PWFA in the blowout regime. In this scheme, up to half a wakefield period is applicable to each proton bunch in terms of long-term and stable deceleration and maintenance. The simulations verify that up to 90% of TeV protons from ten proton bunches survive after propagating through a plasma of length 150 m, bringing the witness electrons to 0.47 TeV with low energy spread (1.3%) and well-preserved normalized emittance. This work expands the concept of proton driven PWFA. Assuming the micro-bunching techniques will be developed in the frame of AWAKE project, the hollow channels would open the path to emittance-preserving accelerating structures for future high-energy physics facilities. The least developed part of the concept is the production of long hollow channels, which is far from maturity[50]. However, the channel could be sectionalized into multiple plasma cells with almost no decrease of the longitudinal fields, provided that the gap between cells is shorter than the betatron period of driver particles[69]. It also can sustain relatively large eigenfrequency fluctuations (at the level of 5%). All of these lower the precision requirements in producing the desired hollow plasma channel.

# Acknowledgements

This work was supported by the President's Doctoral Scholarship Award of The University of Manchester, the Cockcroft Institute core grant and STFC. The authors greatly appreciate the computing time from the clusters at the University of Manchester. Contribution of K. Lotov and A. Sosedkin was supported by SB RAS grant No. 0305-2015-0019.